# First Principles Calculations to Study the Equilibrium Configuration of Ozone Molecule

Laxman Mainali[1,2*], Devendra Raj Mishra[2], Mukunda Mani Aryal[2]

[1]Department of Biophysics, Medical College of Wisconsin, Milwaukee, Wisconsin, USA

[2]Department of Physics, Tribhuvan University, Kirtipur, Nepal

*Author to whom correspondence should be addressed:
Laxman Mainali, Ph.D.
Department of Biophysics
Medical College of Wisconsin
8701 Watertown Plank Road
Milwaukee, WI 53226
Phone: (414) 456-4933
E-mail: lmainali@mcw.edu

**Abstract:** The present work describes the equilibrium configuration of the ozone molecule studied using the Hartree-Fock (HF), Møller-Plesset second order (MP2), Configuration interaction (CI), and Density functional theory (DFT) calculations. With the MP2 calculations, the total energy for the singlet state of ozone molecule has been estimated to be -224.97820 a.u., which is lower than that of the triplet state by 2.01 eV and for the singlet state the bond length and the bond angle have been estimated to be 1.282 Å and 116.85$^0$ respectively. Calculations have also been performed to obtain the total energy of the isomeric excited state of ozone molecule and it is seen that the ground state energy of the ozone molecule is lower than that of the isomeric excited state by 1.79 eV. Furthermore, the binding energy of the ozone molecule has been estimated to be 586.89 kJ/mol. The bond length, bond angle and the binding energy estimated with HF, MP2, CI and DFT are compared with the previously reported experimental values.

## 1 Introduction

Since the first published account of ozone depletion over Antarctica in 1984 by Chubachi of the Japanese Meteorological Research Institute in Ibaraki, the science of ozone depletion has gained remarkable prominence as one of the global environmental issues of the twentieth century [1]. Ozone ($O_3$) is a molecule made up of three oxygen atoms. Averaged over the entire atmosphere, of every 10 million molecules in the atmosphere, only about three are ozone. About 90 percent of ozone is found in the stratosphere, between 10 and 50 kilometres above the earth's surface. This high level ozone plays a crucial role in protecting animals and plants from the sun's harmful ultraviolet (UV) radiation, and stabilizing the earth's climate. This ozone layer is incredibly unstable, since it is constantly being formed and broken down through interactions with UV radiation [2]. The formation and breaking up of ozone molecule can be understood by considering the following reactions [3].

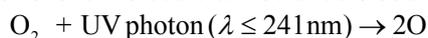

$O_2 + UV\,photon\,(\lambda \leq 241\,nm) \rightarrow 2O$

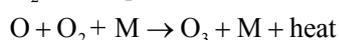

$O + O_2 + M \rightarrow O_3 + M + heat$

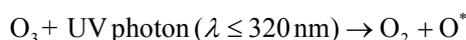

$O_3 + UV\,photon\,(\lambda \leq 320\,nm) \rightarrow O_2 + O^*$

where $\lambda$ is the wavelength of the photon, $M$ is the molecule such as nitrogen molecule ($N_2$) required to carry away heat generated in the collision between atomic oxygen (O) and molecular oxygen ($O_2$) and O* refers to the excited state of atomic oxygen (O). The process of breaking up of ozone molecule in the above reaction by sunlight waves of less than 320 nm is known as non-catalytic destruction of ozone [3].

In addition to the non-catalytic mechanism for the destruction of ozone molecule described above, there are catalytic processes for the destruction of ozone. The catalytic process involves the destruction of ozone molecule by man-made pollutants like chlorofluorocarbons (CFCs), which are chemically inert. CFCs transported by the wind to the stratosphere are broken down by UV radiation, releasing chlorine atoms. Free chlorine atoms released from CFCs molecules react with ozone, forming ClO and $O_2$. ClO is short lived; it reacts with a free O atom to form a further $O_2$ molecule, releasing the free Cl atom. This process becomes a part of chain reaction, as a result of which a single chlorine atom could destroy as many as 100,000 molecules of ozone. This catalytic destruction of ozone molecule causes the depletion of the ozone layer in the stratosphere allowing the sun's harmful UV radiation to reach the earth's surface, endangering human health and the environment, for example by increasing skin cancer and cataracts, weakening human immune systems and damaging crops and natural ecosystems [1]. As animals and plants could not exist without a protective ozone layer in the atmosphere, it is of great importance to understand the processes that regulate the atmosphere's ozone content. With this view in mind, we have carried out ab initio (i.e. first principles) calculations to study the equilibrium configurations for ozone molecule.

## 2. Computational Methods

The first principles calculations have been performed to study the equilibrium configuration of ozone molecule using the Gaussain 98 [4] set of programs. The first principles approaches can be classified into three main categories: the Hartree-Fock approach, the density functional approach and the quantum Monte-Carlo approach [5]. In what follows, we briefly consider the Hartree-Fock method and the density functional theory.

The Hartree-Fock self–consistent method is based on the one-electron approximation in which the motion of each electron in the effective field of all the other electrons is governed by a one-particle Schrodinger equation. The Hartree-Fock approximation takes into account of the correlation arising due to the electrons of the same spin, however, the motion of the electrons of the opposite spin remains uncorrelated in this approximation. The methods beyond self-consistent field methods, which treat the phenomenon associated with the many-electron system properly, are known as the electron correlation methods. One of the approaches to electron correlation is the Møller-Plesset (MP) perturbation theory in which the Hartree-Fock energy is improved by obtaining a perturbation expansion for the correlation energy [6]. However, MP calculations are not variational and can produce an energy value below the true energy [7].

Another approach to electron correlation is the method of configuration interaction (CI), which considers mixing of wave function from the configuration other than the ground state configuration. The full configuration interaction method has many of the desirable features of being well-defined, size-consistent and variational. However, it is almost impractical for all but the very smallest systems. The configuration interaction (CI) method used for many body problems usually augment the Hartree-Fock by adding only limited set of substitutions i.e. truncating the CI expansion at some level of substitutions and CI variants are no longer size consistent [6].

Another first principles approach to calculate the electronic structure for many-electron systems is the density functional theory (DFT). In this theory, the exchange-correlation energy is expressed, at least formally, as a functional of the resulting electron density distribution, and the electronic states are solved for self-consistently as in the Hartree-Fock approximation [5, 8-10]. The density functional theory is, in principle, exact but, in practice, both exchange and dynamic correlation effects are treated approximately [11].

The first principles methods (i.e. HF, HF + MP2, CISD, and DFT) discussed above can be implemented with the aid of the Gaussian-98 set of programs to study the electronic structure and to determine the various physical properties of many-electron systems [12]. Within this framework, we have studied the stability of the ozone molecule using different basis sets. A basis set is the mathematical description of the orbitals within a system (which in turn combine to approximate the total electronic wavefunction) used to perform the theoretical calculation [12]. 3-21G, 3-21G*, 6-31G, 6-31G*, 6-31G**, 6-311G, 6-311G*, 6-311G** are the basis sets used in the calculations. The functional MPW1PW91, which is Barone and Adomo's Becke-style one parameter functional using modified Perdew-Wang exchange and Perdew-Wang 91 correlation (Frisch and Frisch, 1999) [4] is used for DFT Calculations.

## 3. Results and Discussions

The HF, HF+MP2, CISD, and DFT (MPW1PW91) level of calculations for the total energy and the equilibrium geometry of $O_3$ molecule have been performed using the basis sets 3-21G, 6-31G, 6-311G and the corresponding starred sets that allow the inclusion of polarization effects in the wave functions. The result of these calculations has been presented in Tables 1 to 4. The calculated values of the total energy and the optimized geometry for the triplet and singlet states of $O_3$ molecule obtained in different level of approximations with the basis sets mentioned above are the optimized global minimum values except for the HF values of the triplet state of $O_3$ molecule. As it has not been possible to observe the global minimum for the optimized configurations of the triplet state of $O_3$ molecule in the HF level of approximation using the basis sets mentioned above, the HF values for the optimized configurations of the triplet state of $O_3$ molecule presented in Tables 1 and 3 are the local minimum values.

**Table 1 Total energy of triplet state of ozone molecule**

| Basis sets used | Total energy in a.u. calculated in | | | |
|---|---|---|---|---|
| | HF | HF+MP2 | CISD | DFT |
| 3-21G | -223.01421 | -223.39758 | -223.36388 | -224.07063 |
| 3-21G* | -223.01421 | -223.39758 | -223.36388 | -224.07063 |
| 6-31G | -224.15348 | -224.54510 | -224.50386 | -225.23357 |
| 6-31G* | -224.24864 | -224.79432 | -224.73257 | -225.30798 |
| 6-31G** | -224.24864 | -224.79432 | -224.73257 | -225.30798 |
| 6-311G | -224.23004 | -224.64926 | -224.60192 | -225.30886 |
| 6-311G* | -224.30933 | -224.90425 | -224.83642 | -225.36902 |
| 6-311G** | -224.30933 | -224.90425 | -224.83642 | -225.36902 |

**Table 2 Total energy of singlet state of ozone molecule**

| Basis sets used | Total energy in a.u. calculated in | | | |
|---|---|---|---|---|
| | HF | HF+MP2 | CISD | DFT |
| 3-21G | -222.98922 | -223.47124 | -223.38381 | -224.08269 |
| 3-21G* | -222.98922 | -223.47124 | -223.38381 | -224.08269 |
| 6-31G | -224.13863 | -224.61858 | -224.52629 | -225.25083 |
| 6-31G* | -224.26144 | -224.86954 | -224.77283 | -225.34218 |
| 6-31G** | -224.26144 | -224.86954 | -224.77283 | -225.34218 |
| 6-311G | -224.21760 | -224.72198 | -224.62555 | -225.32663 |
| 6-311G* | -224.32264 | -224.97820 | -224.87738 | -225.40291 |
| 6-311G** | -224.32264 | -224.97820 | -224.87738 | -225.40291 |

It is seen from Tables 1 and 2 that the calculated values of total energy for the ozone molecule get lowered with increasing size of the basis sets. The lowering in the energy values on increasing the size of the basis set from 3-21G to 6-31G is found to be around 0.5%, whereas, the corresponding lowering in the energy values on increasing the size of basis set from 6-31G to 6-311G is found to be around 0.03%. With the addition of $d-$type Gaussian polarization functions (i.e. single starred basis sets) to smaller basis sets i.e. 3-21G and 3-21G*, there is no change in the calculated values of energy, whereas, the energy values calculated with the inclusion of $d-$type Gaussian polarization functions to larger basis sets (e.g. 6-31G*, 6-311G*) get lowered as compared to the energy values obtained with the corresponding unstarred basis sets. The lowering in the energy values on changing the basis set from 6-311G to 6-311G* is found to be less than 0.1%. Furthermore, it is seen that the energy values calculated with double starred basis sets (e.g. 6-31G**, 6-311G** which include $d, p-$type Gaussian polarizations functions to 6-31G, 6-311G respectively) do not differ to the energy values obtained with the corresponding single starred basis sets 6-31G* and 6-311G*. From what has been discussed here, it is clear that our results for the total energy of the ozone molecule are basis set convergent.

*Basis set convergence for optimized geometry of O₃ molecule*

Tables 3 and 4 show the values for the optimized geometry of the triplet and singlet states of O$_3$ molecule obtained in the HF, HF+MP2, CISD, and DFT level of approximations using the basis sets mentioned above. It is seen from Tables 3 and 4 that the calculated values of the distance (O-O) for O$_3$ molecule get lowered with increasing size of the basis sets. The lowering in the values of the distance (O-O) on increasing the size of the basis set from 3-21G to 6-31G is found to be around 0.05 Å, whereas, the corresponding lowering in the values of the distance (O-O) on increasing the size of basis set from 6-31G to 6-311G is found to be around 0.01 Å. With the addition of $d-$type Gaussian polarization functions (i.e. single starred basis sets) to smaller basis sets i.e. 3-21G*, there is no change in the calculated values of the distance (O-O), whereas, the values of the distance (O-O) calculated with the inclusion of $d-$type Gaussian polarization functions to larger basis sets (e.g. 6-31G*, 6-311G*) get lowered as compared to the values of the distance (O-O) obtained with the corresponding unstarred basis sets. The lowering in the values of the distance (O-O) on changing the basis set from 6-311G to 6-311G* is found to be around 0.06 Å. Furthermore, it is seen that the values of the distance (O-O) calculated with double starred basis sets (e.g. 6-31G**, 6-311G* which include $d, p-$type Gaussian polarizations functions to 6-31G, 6-311G respectively) do not differ to the values of the distance (O-O) obtained with the corresponding single starred basis sets 6-31G* and 6-311G*.

**Table 3 Optimized geometry of triplet state of ozone molecule**

| Basis sets used | Level of Approximation | | | | | | | |
|---|---|---|---|---|---|---|---|---|
| | HF | | HF+MP2 | | CISD | | DFT (MPW1PW91) | |
| | Bond length (in Å) | Bond Angle (in $\theta^0$) | Bond length (in Å) | Bond angle (in $\theta^0$) | Bond length (in Å) | Bond angle (in $\theta^0$) | Bond length (in Å) | Bond angle (in $\theta^0$) |
| 3-21G | 1.3920 | 127.73 | 1.3434 | 128.30 | 1.3720 | 127.55 | 1.3877 | 126.90 |
| 3-21G* | 1.3920 | 127.73 | 1.3434 | 128.30 | 1.3720 | 127.55 | 1.3877 | 126.90 |
| 6-31G | 1.3176 | 132.72 | 1.3180 | 129.66 | 1.3379 | 130.05 | 1.3423 | 129.62 |
| 6-31G* | 1.2500 | 131.50 | 1.2873 | 129.33 | 1.2783 | 129.97 | 1.2848 | 129.85 |
| 6-31G** | 1.2500 | 131.50 | 1.2873 | 129.33 | 1.2783 | 129.97 | 1.2848 | 129.85 |
| 6-311G | 1.3033 | 132.62 | 1.3130 | 129.69 | 1.3253 | 130.21 | 1.3364 | 129.92 |
| 6-311G* | 1.2409 | 131.30 | 1.2745 | 129.35 | 1.2622 | 129.86 | 1.2788 | 129.91 |
| 6-311G** | 1.2409 | 131.30 | 1.2745 | 129.35 | 1.2622 | 129.86 | 1.2788 | 129.91 |

**Table 4 Optimized geometry of singlet state of ozone molecule**

| Basis sets used | Level of Approximation | | | | | | | |
|---|---|---|---|---|---|---|---|---|
| | HF | | HF+MP2 | | CISD | | DFT (MPW1PW91) | |
| | Bond length (in Å) | Bond angle (in $\theta^0$) | Bond length (in Å) | Bond angle (in $\theta^0$) | Bond length (in Å) | Bond angle (in $\theta^0$) | Bond length (in Å) | Bond angle (in $\theta^0$) |
| 3-21G   | 1.3080 | 117.04 | 1.3895 | 113.53 | 1.3534 | 115.51 | 1.3530 | 116.01 |
| 3-21G*  | 1.3080 | 117.04 | 1.3895 | 113.53 | 1.3534 | 115.51 | 1.3530 | 116.01 |
| 6-31G   | 1.2507 | 119.56 | 1.3566 | 115.24 | 1.3072 | 117.66 | 1.3043 | 117.89 |
| 6-31G*  | 1.2043 | 119.01 | 1.3002 | 116.30 | 1.2446 | 117.78 | 1.2491 | 118.04 |
| 6-31G** | 1.2043 | 119.01 | 1.3002 | 116.30 | 1.2446 | 117.78 | 1.2491 | 118.04 |
| 6-311G  | 1.2451 | 119.78 | 1.3487 | 115.67 | 1.2948 | 118.11 | 1.3004 | 118.22 |
| 6-311G* | 1.1944 | 119.22 | 1.2821 | 116.85 | 1.2273 | 118.14 | 1.2415 | 118.33 |
| 6-311G**| 1.1944 | 119.22 | 1.2821 | 116.85 | 1.2273 | 118.14 | 1.2415 | 118.33 |

It is also seen from Tables 3 and 4 that the HF, HF+MP2, CISD, and DFT values of the angle (O-O-O) for $O_3$ molecule obtained using the basis set 3-21G do not differ with the corresponding values of the angle (O-O-O) obtained with the basis set 6-31G by more than 2% and on changing the basis set from 6-31G to 6-311G the corresponding differences in the calculated values of the angle (O-O-O) do not exceed by more than 0.5%. With the addition of $d$-type Gaussian polarization functions (i.e. single starred basis sets) to smaller basis sets i.e. 3-21G*, there is no change in the calculated values of the angle (O-O-O). However, there is a change of around 0.5% in the values of the angle (O-O-O) with the addition of $d$-type Gaussian polarization functions to larger basis sets (i.e. 6-31G*, 6-311G*). Furthermore, it is seen that the values of the angle (O-O-O) calculated with double starred basis sets (e.g. 6-31G**, 6-311G* which include $d, p$-type Gaussian polarizations functions to 6-31G, 6-311G respectively) do not differ to the values of the angle (O-O-O) obtained with the corresponding single starred basis sets 6-31G* and 6-311G*.

From what has been discussed above, it is clear that our results for the equilibrium geometry of $O_3$ molecule are basis set convergent.

*Stability of $O_3$ molecule*

As stated earlier the estimated values of the energy (E), the distance (O-O) i.e. $d$, and the angle (O-O-O) i.e. A for the singlet and triplet states of $O_3$ molecule in the HF+MP2, CISD, and DFT level of approximations using the basis sets mentioned above are the global minimum values, whereas, the estimated values of the energy, the distance (O-O), and the angle (O-O-O) for the triplet state of $O_3$ molecule in the HF level of calculation are the local minimum values. Hence for a discussion of the stability of $O_3$ molecule in the triplet state, we have focused our attention to the HF+MP2, CISD, and DFT values of the optimized configurations for the triplet state of $O_3$ molecule.

We have made a comparison of the HF+MP2, CISD, and DFT values of the total energy of the triplet state of $O_3$ molecule. It is seen from Table 1 that the values of the total energy for the triplet state of $O_3$ molecule in the DFT level of approximation obtained with all the basis sets as mentioned above are lower than the corresponding HF+MP2 values which, in turn, are lower than the corresponding CISD values. However, the HF+MP2, CISD, and DFT values of the total energy for the triplet state of $O_3$ molecule obtained with a given basis set agree to each other within 0.4%.

We have also made a comparison of the values of the distance (O-O) of the triplet state of $O_3$ molecule calculated in different level of approximations. It is seen from Table 3 that the HF+MP2, CISD, and DFT values of the distance (O-O) of the triplet state of $O_3$ molecule obtained with the smaller basis set (i.e. 3-21G, 3-21G*) and the larger unstarred basis sets (i.e. 6-31G, 6-311G) obey the following inequalities:

$$d_{DFT} > d_{CISD} > d_{HF+MP2}$$

whereas, the corresponding values of the distance (O-O) obtained with the larger polarized basis sets 6-31G*, 6-31G**, 6-311G*, and 6-311G** agree to each other within 0.01 Å. However, the difference in the HF+MP2 and DFT values of the bond length obtained using the smaller basis sets (i.e. 3-21G, 3-21G*) is found to be around 0.04 Å, whereas, the difference in the corresponding values of the bond length obtained using the larger unstarred basis sets (i.e. 6-31G, 6-311G) is found to be around 0.02 Å.

It is also seen from Table 3 that the values of the angle (O-O-O) of the triplet state of $O_3$ molecule calculated in different level of approximations (i.e. HF+MP2, CISD, and DFT) with the smaller basis sets (i.e. 3-21G, 3-21G*) obey the following inequalities:

$$A_{HF+MP2} > A_{CISD} > A_{DFT}$$

whereas, the corresponding values of the angle (O-O-O) obtained with the larger basis sets (6-31G, 6-311G and the corresponding starred sets) agree to each other within 0.5%. However, the difference in the HF+MP2 and DFT values of the angle (O-O-O) obtained with the smaller basis sets (i.e. 3-21G, 3-21G*) is found to be around 1%.

We have also made a comparison of the HF, HF+MP2, CISD, and DFT values of the optimized configurations for the singlet state of $O_3$ molecule. It is seen from Table 2 that the values of the total energy for singlet state of $O_3$ molecule in the HF+MP2 level of approximation obtained with all the basis sets as mentioned above are lower than the corresponding CISD values which, in turn, are lower than the corresponding HF values. It is also seen from Table 2 that the DFT values of the total energy for the singlet state of $O_3$ molecule obtained with all the basis sets as mentioned above are lower than the corresponding HF+MP2 values. However, the HF, HF+MP2, CISD, and DFT values of the total energy for the singlet state of $O_3$ molecule obtained with a given basis set agree to each other within 0.5%.

We have also made a comparison of the values of the distance (O-O) of the singlet state of $O_3$ molecule calculated in different level of approximations. It is seen from Table 4 that the values of the distance (O-O) of the singlet state of $O_3$ molecule calculated in different level of approximations with the basis sets 3-21G, 3-21G*, and 6-31G obey the following inequalities:

$$d_{HF+MP2} > d_{CISD} > d_{DFT} > d_{HF}$$

whereas, the corresponding values of the distance (O-O) obtained with the basis sets 6-311G, 6-31G*, 6-31G**, 6-311G*, and 6-311G** obey the following relation

$$d_{HF+MP2} > d_{DFT} > d_{CISD} > d_{HF}$$

However, the differences in the HF and HF+MP2 values of the distance (O-O) obtained with the basis sets mentioned above is found to be around 0.1 Å.

Furthermore, a comparison of the values of the angle (O-O-O) of singlet state of $O_3$ molecule (Table 4) calculated in different level of approximations with the basis sets (3-21G, 6-31G, 6-311G and the corresponding starred sets) used shows the values of the bond angle obtained in different level of approximations to obey the following inequalities:

$$A_{HF} > A_{DFT} > A_{CISD} > A_{HF+MP2}$$

However, the difference in the HF and HF+MP2 values of the angle (O-O-O) obtained with the basis sets mentioned above is found to be around 4%.

We have also made a comparison between the values of the optimized geometry of the singlet and triplet states of ozone molecule obtained in the HF+MP2, CISD, and DFT level of approximations using the basis sets mentioned above. It is also seen from Tables 3 and 4 that the HF+MP2 values of the distance (O-O) for the triplet state of $O_3$ molecule obtained with the basis sets (i.e. 3-21G, 6-31G, 6-311G and the corresponding starred sets) are lower as compared to the corresponding values of the distance (O-O) in the singlet state. The lowering in the values of the distance (O-O) using the basis sets (3-21G, 3-21G*, 6-31G, 6-311G) is found to be around 0.04 Å, whereas, the lowering in the values of the distance (O-O) using the basis sets (6-31G*, 6-31G**, 6-311G*, and 6-311G**) is found to be around 0.01 Å. It is also seen from Tables 3 and 4 that the CI and DFT values of the distance (O-O) for the singlet state of $O_3$ molecule obtained with the basis sets (i.e. 3-21G, 6-31G, 6-311G and the corresponding starred sets) are lower as compared to the corresponding values of the distance (O-O) in the triplet state. The lowering in the values of the distance (O-O) in the CISD level of calculation is found to be around 0.03 Å, whereas, the lowering in the values of the distance (O-O) in the DFT level of calculation is found to be around 0.04 Å.

Furthermore, it is also seen from Tables 3 and 4 that the HF+MP2, CISD, and DFT values of the angle (O-O-O) for the singlet state of $O_3$ molecule obtained with the basis sets (i.e. 3-21G, 6-31G, 6-311G and the corresponding starred sets) are lower as compared to the corresponding values of the angle (O-O-O) in the triplet state. The lowering in the values of the angle (O-O-O) in the HF+MP2 level of calculation is found to be around $14^0$, whereas, the lowering in the values of the angle (O-O-O) in the CISD and DFT level of calculations is found to be around $12^0$.

A close look at Tables 1 and 2 shows that the energy values of the singlet state of $O_3$ molecule obtained in the HF+MP2, CISD, and DFT level of approximations with all the basis sets (3-21G, 6-31G, 6-311G and the corresponding starred sets) used are lower than that of the triplet state indicating that the singlet state of $O_3$ molecule is a stable state. The difference in the energy between the triplet and singlet states (i.e. $\delta E = E_{triplet} - E_{singlet}$) of $O_3$ molecule calculated in the HF+MP2, CISD, and DFT level of calculations with all the basis sets (3-21G, 6-31G, 6-311G and the corresponding starred sets) have been presented in Table 5.

**Table 5 Energy difference between the triplet and singlet states of ozone molecule**

| Basis sets used | Energy difference (i.e. $\delta E = E_{triplet} - E_{singlet}$) in eV calculated in | | |
|---|---|---|---|
| | HF+MP2 | CISD | DFT |
| 3-21G | 2.00 | 0.54 | 0.33 |
| 3-21G* | 2.00 | 0.54 | 0.33 |
| 6-31G | 2.00 | 0.61 | 0.47 |
| 6-31G* | 2.05 | 1.10 | 0.93 |
| 6-31G** | 2.05 | 1.10 | 0.93 |
| 6-311G | 1.98 | 0.64 | 0.48 |
| 6-311G* | 2.01 | 1.11 | 0.92 |
| 6-311G** | 2.01 | 1.11 | 0.92 |

The HF+MP2, CISD, and DFT values of the energy difference ($\delta E$) between the triplet and singlet states of $O_3$ molecule calculated in different level of approximations obtained with all the basis sets mentioned above show the following feature i.e.

$$\delta E_{HF+MP2} > \delta E_{CISD} > \delta E_{DFT}$$

It is seen from Table 5 that the HF+MP2 values of $\delta E$ obtained with the basis sets mentioned above are close to each other within 3%. However, the CISD and the DFT values of $\delta E$ obtained with the unstarred basis sets (i.e. 3-21G, 6-31G, 6-311G) and smaller starred basis set (i.e. 3-21G*) are considerably lower than the corresponding values of $\delta E$ obtained with the larger starred basis sets (i.e. 6-31G*, 6-31G**, 6-311G*, 6-311G**). As it is generally expected that the basis set of higher flexibility would be a better approximation, the energy difference (i.e. $\delta E = E_{triplet} - E_{singlet}$) between the triplet and singlet states of $O_3$ molecule has been estimated in the HF+MP2, CISD, and DFT level of approximations using the basis set 6-311G**. Our estimated values for the energy difference (i.e. $\delta E = E_{triplet} - E_{singlet}$) between the triplet and singlet states of $O_3$ molecule are found to be 2.01, 1.11, and 0.92 eV in the HF+MP2, CISD, and DFT level of approximations respectively. These variations in the values of $\delta E$ obtained in different level of approximations could be due to the loss of significant figures in subtracting two large numbers of almost equal magnitude.

We have also studied the variation of the total energy of the triplet and singlet states of $O_3$ molecule with respect to the bond length (O-O) and the bond angle (O-O-O) obtained in the HF+MP2, CISD, and DFT level of approximations using the basis set 6-311G**. The curves showing the variation of energy with the bond length (O-O) and the bond angle (O-O-O) for the triplet and singlet states of $O_3$ molecule with HF+MP2 level of calculations have been shown in Figs. 1 and 2. Similar trend follows for CISD and DFT calculations (data not shown). It is seen from Figs. 1 and 2 that the energy minimum occurs for both the singlet and triplet states of $O_3$ molecule. The HF+MP2, CISD, and DFT values of the minimum energy for the triplet state of $O_3$ molecule are found to be -224.904, -224.836, and -224.369 a.u. respectively, whereas, the corresponding values of the minimum energy for the singlet state of $O_3$ molecule are found to be -224.978, -224.877, and -224.403 a.u. respectively. The HF+MP2, CISD, and DFT values of the equilibrium bond length (O-O) and the bond angle (O-O-O) for the triplet state of $O_3$ molecule are found to be 1.27, 1.26, 1.28 Å and $129^0$, $130^0$, $128^0$ respectively, whereas, the corresponding values of the equilibrium bond length (O-O) and the bond angle (O-O-O) for the singlet state of $O_3$ molecule are found to be 1.28, 1.23, 1.24 Å and $117^0$, $118^0$, $118^0$ respectively.

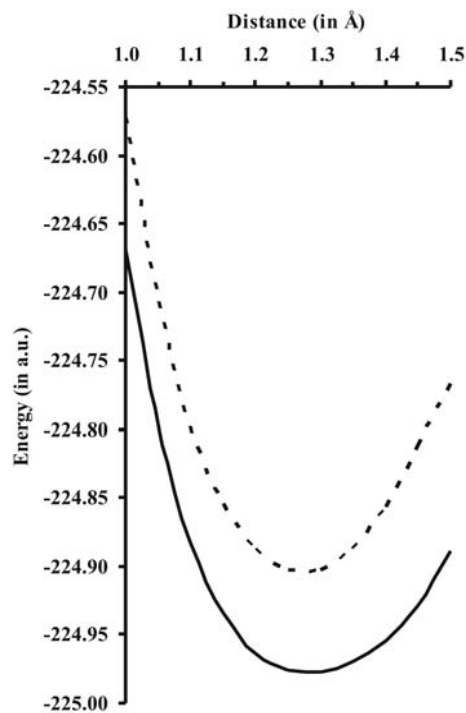

**Figure 1: Variation of the HF+MP2 energy of the triplet (dotted line) and singlet state (solid line) with bond length (O-O) at the bond angle of $129.35^0$ (triplet state) and $116.85^0$ (singlet state) using the basis set 6-311G\*\***

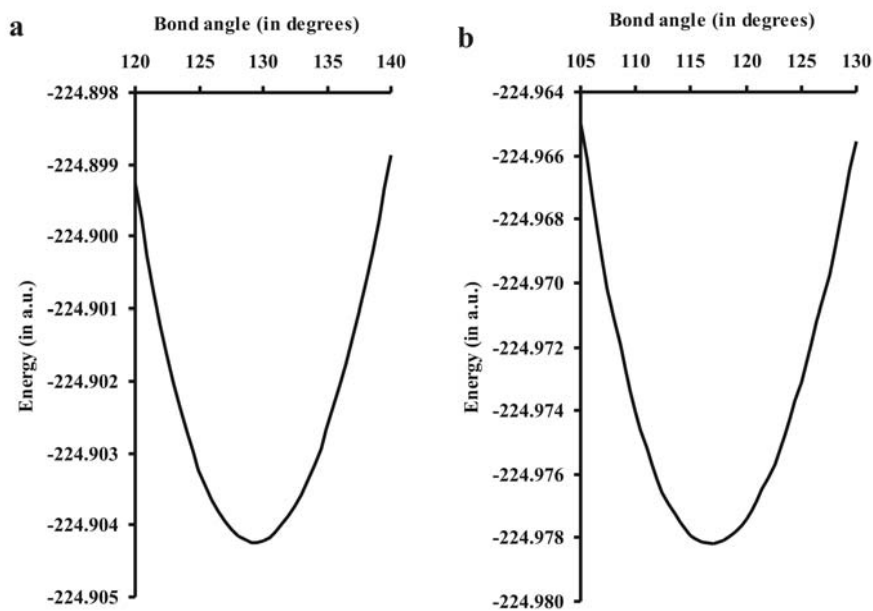

**Figure 2: Variation of the HF+MP2 energy of the triplet (a) and singlet state (b) with bond angle (O-O-O) at the bond length of 1.2745 Å (triplet state) and 1.2821 Å (singlet state) using the basis set 6-311G\*\*.**

From what has been discussed above, it is clearly seen that in all the level of approximations (i.e. HF+MP2, CISD, and DFT) studied the singlet state energy is found to be lower than that of the triplet state for $O_3$ molecule. With this analysis, we present an estimate of the equilibrium configuration of the singlet state of $O_3$ molecule obtained in the HF+MP2 level of approximation using the basis set 6-311G\*\*. Our estimated values of the bond length (O-O), bond angle (O-O-O), and the ground state energy of $O_3$ molecule are 1.282 Å, $116.85^0$, and -224.97820 a.u. respectively and the equilibrium configuration of the ground state of $O_3$ molecule is as given in Fig 3.

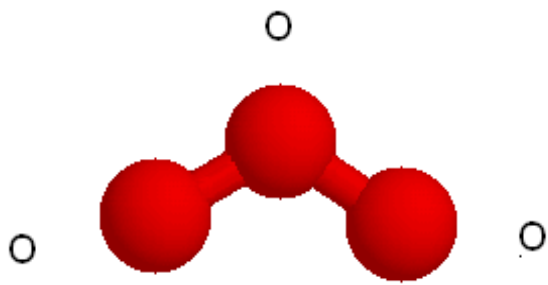

**Figure 3 A ground state of ozone molecule at the configuration of global minimum**

*Isomeric excited state of $O_3$ molecule*

We have also calculated the total energy and the bond distance (O-O) of the isomeric excited state of $O_3$ molecule where O atoms occupy three corners of an equilateral triangle. The total energy and the bond distance (O-O) of the isomeric excited state of $O_3$ molecule obtained in the HF, HF+MP2, CISD, and DFT level of approximations using the basis sets 3-21G, 6-31G, 6-311G and the corresponding starred sets which allow the inclusion of polarization effects in the wave functions have been presented in Table 6. It is seen from Table 6 that the calculated values of the equilibrium configuration for the isomeric excited state of $O_3$ molecule obtained in the HF, HF+MP2, CISD, and DFT (MPW1PW91) level of approximations using the basis sets mentioned above show similar basis set convergence as that of $O_3$ molecule in the ground state.

**Table 6 Total energy and bond distance of isomeric excited state of ozone molecule**

| Basis sets used | Level of Approximation | | | | | | | |
|---|---|---|---|---|---|---|---|---|
| | HF | | HF+MP2 | | CISD | | DFT (MPW1PW91) | |
| | Energy (in a.u.) | Distance (in Å) | Energy (in a.u.) | Distance (in Å) | Energy (in a.u.) | Distance (in Å) | Energy (in a.u.) | Distance (in Å) |
| 3-21G | -223.02616 | 1.4867 | -223.45910 | 1.5989 | -223.40000 | 1.5388 | -224.07802 | 1.5262 |
| 3-21G* | -223.02616 | 1.4867 | -223.45910 | 1.5989 | -223.40000 | 1.5388 | -224.07802 | 1.5262 |
| 6-31G | -224.12580 | 1.4668 | -224.58415 | 1.6394 | -224.50611 | 1.5400 | -225.21264 | 1.5162 |
| 6-31G* | -224.24479 | 1.3727 | -224.81350 | 1.4777 | -224.74264 | 1.4199 | -225.29950 | 1.4178 |
| 6-31G** | -224.24479 | 1.3727 | -224.81350 | 1.4777 | -224.74264 | 1.4199 | -225.29950 | 1.4178 |
| 6-311G | -224.20100 | 1.4366 | -224.67934 | 1.6264 | -224.59804 | 1.5126 | -225.28517 | 1.5004 |
| 6-311G* | -224.30306 | 1.3552 | -224.91236 | 1.4493 | -224.84147 | 1.3928 | -225.35498 | 1.4068 |
| 6-311G** | -224.30306 | 1.3552 | -224.91236 | 1.4493 | -224.84147 | 1.3928 | -225.35498 | 1.4068 |

We have also made a comparison of the HF, HF+MP2, CISD, and DFT values of the optimized configurations for the isomeric excited state of $O_3$ molecule. It is seen from Table 6 that the values of the total energy for isomeric excited state of $O_3$ molecule in the HF+MP2 level of approximation obtained with all the basis sets as mentioned above are lower than the corresponding CISD values which, in turn, are lower than the corresponding HF values. Furthermore, it is also seen from Table 6 that the DFT values of the total energy for isomeric excited state of $O_3$ molecule obtained with all the basis sets as mentioned above are lower than the corresponding HF+MP2 values. However, the HF, HF+MP2, CISD, and DFT values of the total energy for the isomeric excited state of $O_3$ molecule obtained with a given basis set agree to each other within 0.5%.

We have also made a comparison of the values of the distance (O-O) of the isomeric excited state of $O_3$ molecule calculated in different level of approximations. It is seen form Table 6 that the values of the distance (O-O) of the isomeric excited state of $O_3$ molecule calculated in different level of approximations with the basis sets 3-21G, 3-21G*, 6-31G, 6-31G*, 6-31G**, 6-311G obey the following inequalities:

$$d_{HF+MP2} > d_{CISD} > d_{DFT} > d_{HF}$$

whereas, the corresponding values of the distance (O-O) obtained with the basis sets 6-311G* and 6-311G** obey the following inequalities:

$$d_{HF+MP2} > d_{DFT} > d_{CISD} > d_{HF}$$

However, the difference in the HF and HF+MP2 values of the distance (O-O) obtained using the basis sets (i.e. 3-21G, 3-21G*, 6-31G*, 6-31G**, 6-311G*, 6-311G**) is found to be around 0.1 Å, whereas the difference in the corresponding values of the distance (O-O) obtained using the basis sets (i.e. 6-31G, 6-311G) is found to be around 0.2 Å.

As it is generally expected, using the basis set of higher flexibility would be a better approximation; we present an estimate of the optimized configuration for the isomeric excited state of $O_3$ molecule obtained in the HF+MP2 level of approximation using the basis set 6-311G**. Our estimated values of the bond distance (O-O) and the total energy of $O_3$ molecule are 1.449 Å and -224.91236 a.u. respectively and the equilibrium configuration of the isomeric excited state of $O_3$ molecule is as given in Fig. 4.

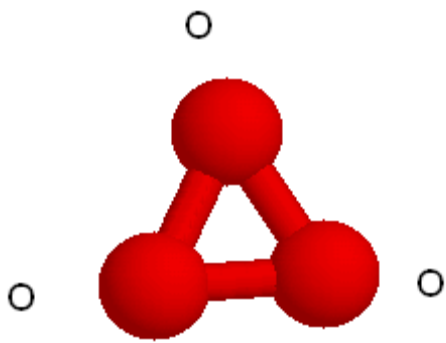

**Figure 4 An isomeric excited state of ozone molecule at the configuration of global minimum**

*Equilibrium Configuration of $O_3$ molecule*

It is seen from Figs 3 and 4 that the difference in the value of the bond length (O-O) between the ground state of $O_3$ molecule and the isomeric excited state of $O_3$ molecule is found to be around 0.17 Å.

We have also calculated the energy difference ($\delta E$) between the minimum energy values of the equilibrium configurations of $O_3$ molecule as given in Figs. 3 and 4. The estimated values of the energy difference ($\delta E$) obtained in the HF, HF+MP2, CISD, and DFT level of calculations with all the basis sets (3-21G, 6-31G, 6-311G and the corresponding starred sets) have been presented in Table 7.

**Table 7 Energy difference between the minimum energy values of the equilibrium configurations of $O_3$ molecule as given in Figs. 3 and 4.**

| Basis sets used | Energy difference ($\delta E$) in eV calculated in | | | |
|---|---|---|---|---|
| | HF | HF+MP2 | CISD | DFT |
| 3-21G | -1.00 | 0.33 | -0.44 | 0.13 |
| 3-21G* | -1.00 | 0.33 | -0.44 | 0.13 |
| 6-31G | 0.35 | 0.94 | 0.55 | 1.04 |
| 6-31G* | 0.45 | 1.52 | 0.82 | 1.16 |
| 6-31G** | 0.45 | 1.52 | 0.82 | 1.16 |
| 6-311G | 0.45 | 1.16 | 0.75 | 1.13 |
| 6-311G* | 0.53 | 1.79 | 0.98 | 1.30 |
| 6-311G** | 0.53 | 1.79 | 0.98 | 1.30 |

It is seen from Table 7 that the HF+MP2 and DFT values of $\delta E$ obtained with all the basis sets (3-21G, 6-31G, 6-311G and the corresponding starred sets) used and the HF and CISD values obtained with the basis sets (6-31G, 6-311G and the corresponding starred sets) are positive. However, the HF and CISD values of $\delta E$ obtained with the smaller basis sets (i.e. 3-21G, 3-21G*) are negative. In what follows, we do not consider the negative values of $\delta E$ obtained with the HF and CISD level of calculation using the smaller basis sets (i.e. 3-21G, 3-21G*) as the HF+MP2 and DFT values of $\delta E$ obtained with all the basis sets and the HF and CISD values obtained with the basis sets of higher flexibility are positive. These observations indicate that the equilibrium configuration of $O_3$ molecule as given in Fig. 3 has lower energy than that of the equilibrium configuration as given in Fig. 4. As the estimation of $\delta E$ involves the difference between two large numbers of almost equal magnitude the variations in the values of $\delta E$ obtained in different level of approximations could be due to the loss of significant figures in the process of subtraction.

We have also studied the variation of the total energy of the isomeric excited state of $O_3$ molecule with respect to the distance between two oxygen atoms (O-O) obtained in the HF, HF+MP2, CISD, and DFT level of approximations using the basis set 6-311G**. The curves showing the variation of the HF+MP2 energy with the distance (O-O) in the range 1.2 Å to 1.8 Å for the isomeric excited state of $O_3$ molecule have been shown in Figure 5. Similar trend follows for HF, CISD and DFT calculations (data not shown). On varying the distance ($d$) between the two oxygen atoms of $O_3$ molecule in the isomeric excited state, the energy minima occurs at 1.36, 1.45, 1.39, and 1.41 Å in the HF, HF+MP2, CISD, and DFT level of approximations respectively. The corresponding values of the minimum energy are found to be -224.30, -224.91, -224.84, and -224.35 a.u. respectively.

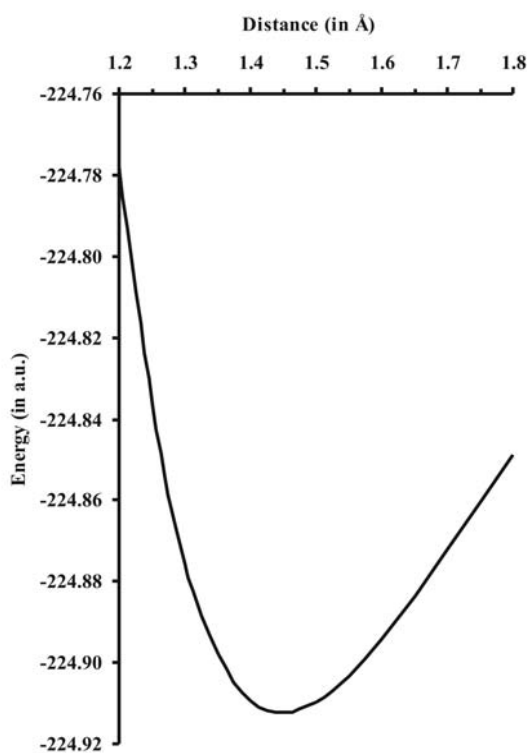

**Figure 5. Variation of the HF+MP2 energy of the isomeric excited state of ozone molecule with bond length (O-O) at the bond angle of $60^0$ using the basis set 6-311G\*\*.**

From the comparison of energies at two different equilibrium configuration (Figs. 3 and 4), it is found that the HF, HF+MP2, CISD, and DFT values of the minimum energy for the equilibrium configuration of $O_3$ molecule (Fig. 3) are lower than the corresponding values of the minimum energy of the equilibrium configuration of $O_3$ molecule (Fig. 4) by 0.020, 0.068, 0.038, 0.053 a.u. respectively. From what has been discussed above it is clear that the equilibrium configuration of $O_3$ molecule as described in Fig. 3 is the minimum energy configuration as compared to the equilibrium configuration as described in Fig. 4.

*Binding energy of $O_3$ molecule*

We present an estimate for the binding energy of $O_3$ molecule considering the equilibrium configuration as in Fig. 4.22. We have estimated the binding energy ( B.E ) of $O_3$ molecule with the aid of the following relation:

$B.E = 3 E(O) - E(O_3)$

where $E(O)$ and $E(O_3)$ denote the ground state energy of O atom and $O_3$ molecule respectively. The results of these calculations have been presented in Table 8. Table 8 shows the values of the binding energy for $O_3$ molecule calculated in the HF, HF+MP2, CISD, and DFT level of approximations using the basis sets 3-21G, 6-31G, 6-311G and the corresponding starred sets that allow the inclusion of polarization effects in the wave functions.

**Table 8 Binding energy of ozone molecule**

| Basis sets used | Binding energy in kcal/mol calculated in | | | |
|---|---|---|---|---|
| | HF | HF+MP2 | CISD | DFT |
| 3-21G | -120.327 | 88.618 | 22.898 | 94.593 |
| 3-21G* | -120.327 | 80.618 | 22.898 | 94.593 |
| 6-31G | -126.947 | 80.567 | 8.836 | 81.248 |
| 6-31G* | -56.704 | 143.972 | 56.852 | 133.931 |
| 6-31G** | -56.704 | 143.972 | 56.852 | 133.931 |
| 6-311G | -119.156 | 87.116 | 12.723 | 83.058 |
| 6-311G* | -58.354 | 140.411 | 53.315 | 128.263 |
| 6-311G** | -58.354 | 140.411 | 53.315 | 128.263 |

It is seen from Table 8 that the HF values of the binding energy for O$_3$ molecule obtained using the basis sets (3-21G, 6-31G, 6-311G and the corresponding starred sets) are negative, indicating that there is no binding. However, it is experimentally known that O$_3$ molecule is stable with a binding energy of 142.2 kcal/mol [12]. This clearly shows that the one-electron HF level of calculations cannot account for the binding of O$_3$ molecule.

With the Møller-Plesset second order perturbation (MP2) calculations, it is seen from Table 8 that the values of the binding energy for O$_3$ molecule are positive with all the classes of the basis sets used. As the HF+MP2 calculation takes into account of electron correlation effects, the binding of O$_3$ molecule can be considered arising from the contribution of electron correlation to the equilibrium configurations of O$_3$ molecule. The HF+MP2 values of the binding energy for O$_3$ molecule obtained with the smaller starred and unstarred basis sets (i.e. 3-21G and 3-21G*) are same, whereas, the HF+MP2 values of the binding energy obtained with the larger starred basis set (i.e. 6-311G* which include $d-$type Gaussian polarization functions) are larger by around 38% than the values obtained with the corresponding unstarred basis sets (i.e. 6-311G). Furthermore, it is seen that the HF+MP2 values of the binding energy calculated with the double starred basis sets (e.g. 6-31G**, 6-311G** which include $d, p-$type Gaussian polarization functions to 6-31G, 6-311G respectively) do not differ to the HF+MP2 values of the binding energy obtained with the corresponding single starred basis sets 6-31G* and 6-311G*. Furthermore, the HF+MP2 values of the binding energy for O$_3$ molecule obtained with the basis sets including $d-$ and $d, p-$type Gaussian polarization functions are close to the experimental value.

It is also seen from Table 8 that the values of the binding energy for O$_3$ molecule obtained with the CISD and DFT level of approximations show similar basis set dependence as of the corresponding HF+MP2 values. With this analysis, we present an estimate for the binding energy and the bond length of O$_3$ molecule using the basis set 6-311G** which include $d, p-$type Gaussian polarization functions to 6-311G. Our estimated values of the binding energy and the bond length for O$_3$ molecule obtained in the HF+MP2, CISD, and DFT level of approximations with the basis set 6-311G** along with the corresponding experimental values have been presented in Table 9.

**Table 9** Comparison between the estimated and the experimentally observed values of the bond length, the bond angle, and the binding energy for O$_3$ molecule in the ground state

| Parameters | Level of calculation | Estimated values | | Experimental values [b] |
|---|---|---|---|---|
| Bond length (in Å) | HF+MP2 | 1.282 | (<1%)[a] | 1.272 |
| | CISD | 1.227 | (4%) | |
| | DFT | 1.242 | (2%) | |
| Bond angle (in degree) | HF+MP2 | 116.85 | (<0.1%) | 116.8 |
| | CISD | 118.14 | (1%) | |
| | DFT | 118.33 | (1%) | |
| Binding energy (in kcal/mol) | HF+MP2 | 140.41 | (1%) | 142.2 |
| | CISD | 53.31 | (62%) | |
| | DFT | 128.26 | (10%) | |

[a] The values given in parentheses denote the percentage deviation in the estimated values of the bond length, the bond angle and the binding energy from the corresponding experimental values.
[b] The experimental values are from Foresman and Frisch [12].

It is seen from Table 9 that the HF+MP2 value of the bond length for O$_3$ molecule is larger than the DFT value by around 3% which, in turn, is larger than the CISD value by around 1%. However, the HF+MP2, CISD and DFT values of the bond length for O$_3$ molecule agree to each other within 4%. The HF+MP2 value of 1.282 Å for the bond length (O-O) of O$_3$ molecule is close to the experimental value of 1.272 Å within 1%.

It is also seen from Table 9 that the DFT values of the bond angle (O-O-O) for O$_3$ molecule is close to the CISD value within 0.2% and is larger than the HF+MP2 value by around 1%. The HF+MP2 value of 116.85$^0$ for the bond angle (O-O-O) of O$_3$ molecule is close to the experimental value of 116.8$^0$ within 0.05%.

Furthermore, it is seen from Table 9 that the values of the binding energy for O$_3$ molecule calculated using the CISD level of approximation is far far less than the DFT value by around 58%, which, in turn, is smaller by around 9% than the HF+MP2 value. The HF+MP2 value of 140.41 kcal/mol for the binding energy of O$_3$ molecule is close to the experimental value of 142.2 kcal/mol within 1%.

## 4. Concluding Remarks

We have studied the equilibrium configurations for the ozone molecule in the HF, HF+MP2, CISD, and DFT level of approximations using the basis set 3-21G, 6-31G, 6-311G and the corresponding starred sets that allow the inclusion of polarization effects in the wave functions. The consistency of the results obtained is tested by their convergence with respect to the use of basis sets of increasing size and complexity and it is seen that our results for the equilibrium configurations of the ozone molecule are basis set convergent. As the use of the basis set of higher flexibility would provide a better estimation, in what follows, we present an estimate of the equilibrium configurations for ozone molecule obtained in the HF+MP2 level of approximation using the basis set 6-311G**.

We have studied the equilibrium configuration for the ozone molecule ($O_3$) in the HF+MP2 level of approximation using the basis set 6-311G**. Our estimated values of the total energy for the triplet and singlet states of the ozone molecule are -224.90425 and -224.97820 a.u. respectively. It is seen that the energy of the singlet state of the ozone molecule is lower than that of the triplet state by 2.01 eV. We have estimated the bond length, bond angle, and the binding energy of $O_3$ molecule to be 1.282 Å, $116.85^0$, and 141.1 kcal/mol respectively, which agree well with the corresponding experimental values of 1.272 Å, $116.8^0$, and 142.2 kcal/mol [12]. We have also studied the isomeric excited state of $O_3$ molecule where O atoms occupy three corners of an equilateral triangle as in Fig. 4. The HF+MP2 values of the energy and the bond length for the equilibrium configuration of $O_3$ as given in Fig. 3 are lower than the corresponding values of the energy and the bond length of the equilibrium configuration of $O_3$ molecule as given in Fig. 4. by 1.79 eV and 0.17 Å respectively.

From what has been discussed above, it is clearly seen that the bond length, bond angle, and the binding energy obtained for the ozone molecule using the first-principles calculations are in close agreement with the corresponding experimental values [12]. This close agreement between the calculated and the experimental values for the equilibrium configurations of the ozone molecule suggests that the first-principles calculations for the excited states of the ozone molecule and its clusters would be illuminating for a better understanding of the dissociation process of $O_3$ molecule.

## Acknowledgements

We are grateful to Prof. T. P. Das (University at Albany, State University of New York) for providing Gaussian 98 program for computation.